\documentclass[a4paper]{jpconf}
\usepackage{graphicx}
\begin{document}
\title{Reduced Hamiltonian for intersecting shells and Hawking radiation}

\author{Pietro Menotti}

\address{Department of Physics, University of Pisa, Italy and
INFN Sezione di Pisa, Italy}

\ead{menotti@df.unipi.it}

\begin{abstract}
We consider the dynamics of one or more self gravitating shells of matter in
  a centrally symmetric gravitational field in the Painlev\'e family of
  gauges. We give the reduced hamiltonian for two intersecting shells, both
  massless and massive. Such a formulation is applied to the computation of
  the semiclassical action of two intersecting shells. The relation of the
  imaginary part of the space-part of the action to the computation of the
  Bogoliubov coefficients is revisited.

\end{abstract}

\section{Introduction}\label{introSec}

In this paper we shall study the dynamics of one or more self-gravitating 
spherical shells
of matter subject to a centrally symmetric gravitational field and the
application of the ensuing formalism to the semiclassical treatment of Hawking
radiation. Such field of research was started by the papers by Kraus and
Wilczek \cite{KW1,KW2}. The main results we shall report here are the
extension of the treatment to one massive shell of matter and also the
extension to two or more massive shells of matter which in the time
development can also intersect \cite{FM}. The main appeal of the approach of 
\cite{KW1,KW2} is that energy conservation is taken exactly into account 
and thus the back reaction effects can be computed.

We shall keep the formalism for more that one shell as close as possible
to the original formalism of \cite{KW1} which can be summarized in words that
we shall adopt a Painlev\'e gauge. 

The extraction of the reduced hamiltonian is often classified as a very
complicated procedure. We shall show that by introducing a ``generating
function'' the derivation can be drastically simplified and also by the same
token it can be extended to deal not only with massive shells but also with a
finite number of massive shells. The treatment of massless shells
is just a particular case.

In Sec.\ref{reducedSec} we shall give the derivation of the reduced 
action in the case of one
shell while proving the independence of the canonical momentum within the
Painlev\'e class of gauges. In Sec.\ref{eqmotionSec} we shall discuss the 
equations of motion.
In Sec.\ref{analyticSec}  we work out the analytic properties of the 
conjugate momentum $p_c$ which
appears in the reduced action and in Sec.\ref{twoshellSec} we shall briefly 
describe the
extension of the treatment to more than one shell. An application of the
results is given in Sec.\ref{exchangeSec}  giving a derivation within such 
a formalism of
the Dray-'t Hooft- Redmount exchange relations. An important integrability
relation is reported in  Sec.\ref{integrabilitySec}  which allows to compute 
the imaginary part of
the space integral of the canonical momenta for two shells. In
Sec.\ref{modeSec} we
revisit the role of the imaginary part of the canonical momentum in
determining the Bogoliubov coefficients and thus the most important features of
Hawking radiation. In Sec.\ref{conclusionSec} we give some concluding remarks.

\section{The reduced action}\label{reducedSec}
As usual we write the metric for a spherically symmetric configuration
in the ADM form 
\begin{equation}
ds^2=-N^2 dt^2+L^2(dr+N^r dt)^2+R^2d\Omega^2.
\end{equation}
were following \cite{KW1,KW2,FLW,LWF} we shall choose the functions $N, N^r, L,
R$ as continuous functions of the coordinates. We shall work on a finite 
region of space time $(t_i, t_f)\times( r_0,
r_m)$ . On the two initial and final surfaces we give the intrinsic
metric by specifying $R(r,t_i)$ and $L(r,t_i)$ and similarly
$R(r,t_f)$ and $L(r,t_f)$.

\bigskip

\noindent The complete action in hamiltonian form, boundary terms included is
\cite{KW1,KW2,hawkinghunter,FMP} 
\begin{eqnarray}\label{completeaction}
S=S_{shell}+\int_{t_i}^{t_f} dt 
\int_{r_0}^{r_m} dr (\pi_L \dot L +
\pi_R \dot R - N{\cal H}_t-N^r {\cal H}_r) + \nonumber \\
\int_{t_i}^{t_f} dt \left.(-N^r \pi_L L+ \frac{NRR'}{L})\right|^{r_m}_{r_0}
\end{eqnarray}
where
\begin{equation}\label{shellaction}
S_{shell}=\int_{t_i}^{t_f} dt ~\hat p~\dot{\hat r}. 
\end{equation}
$\hat r$ is the shell position and $\hat p$ its canonical conjugate
momentum. ${\cal H}_r$ and ${\cal H}_t$ are the constraints
\begin{equation}\label{constraintr}
{\cal H}_r = \pi_R R'- \pi_L'L -\hat p~\delta(r-\hat r),
\end{equation}
\begin{equation}\label{constraintt}
{\cal H}_t = \frac{R R''}{L}+\frac{{R'}^2}{2
L}+\frac{L \pi_L^2}{2R^2}-\frac{R R' L'}{L^2}-
\frac{\pi_L\pi_R}{R}-\frac{L}{2}+ \sqrt{{\hat p}^2
L^{-2}+m^2}~\delta(r-\hat r).
\end{equation}
Action (\ref{completeaction}) is immediately generalized to a finite
number of shells.  
The shell action as given by 
eqs.(\ref{shellaction},\ref{constraintr},\ref{constraintt}) refers to a dust
shell even 
though generalizations to more complicated equations of state have been
considered \cite{FMP,goncalves,nunez}. The boundary terms are those
given in the paper by Hawking and Hunter \cite{hawkinghunter} and will play a
very important role in the following. The function $F$ \cite{FM}
\begin{equation}\label{Fgeneral}
F=R L \sqrt{\left(\frac{R'}{L}\right)^2 -1+ \frac{2{\cal M}}{R}} +
RR'\log\left(\frac{R'}{L}- \sqrt{\left(\frac{R'}{L}\right)^2 -1+ 
\frac{2{\cal M}}{R}}\right)
\end{equation}
has the remarkable property of generating the conjugate momenta as solutions
of the constraints as follows
\begin{equation}\label{piLfromF}
\pi_L =\frac{\delta F}{\delta L}=\frac{\partial F}{\partial L}
\end{equation}
\begin{equation}\label{piRfromF}
\pi_R = \frac{\delta F}{\delta R}=\frac{\partial F}{\partial
R}-\frac{\partial }{\partial r}\frac{\partial F}{\partial R'} 
\end{equation}
where ${\cal M}$ is a mass which is constant in $r$ except at the shell
position $\hat r$ and that we shall denote by $H$ for $r$ above all the shell
positions and by $M$ for $r$ below all the shell positions.
We shall adopt a Painlev\'e gauge defined by $L=1$ everywhere. With regard
to the remaining freedom in the choice of the gauge we shall choose $R=r$
except for a deformation region near the shell positions. Such deformation is
unavoidable because the constraints impose a discontinuity in the derivative
$R'(r)$ at $r=\hat r$ as we shall see in the following.
In the Painlev\'e gauges $F$ becomes
\begin{equation}
F = R W(R,R',{\cal M}) +
RR'({\cal L}(R,R',{\cal M}) - {\cal B}(R,{\cal M}))
\end{equation}
where
\begin{equation}\label{Wdefinition}
W(R,R',{\cal M})= \sqrt{R'^2-1+\frac{2{\cal M}}{R}};~~
{\cal L}(R,R',{\cal M}) = \log(R'-W(R,R',{\cal M}))
\end{equation}
and
\begin{equation}
{\cal B}(R,{\cal M}) = \sqrt{\frac{2{\cal M}}{R}}+
\log\left(1-\sqrt{\frac{2{\cal M}}{R}}\right)
\end{equation}
where we exploited the freedom of adding to (\ref{Fgeneral}) a total 
derivative, thus
gaining for $F$ the useful property of vanishing wherever $R'=1$.
For $L=1$ eqs.(\ref{piLfromF},\ref{piRfromF}) become
\begin{equation}
\pi_L=R\sqrt{R'^2-1+\frac {2{\cal M}}{R}}\equiv R W;~~~~
\pi_R= \frac{[R R''+R'^2-1+{\cal M}/R]}{W}.
\end{equation}

\noindent With regard to $R(r,t)$ one can choose several gauges, within the
Painlev\'e family.

\begin{figure}[h]
\begin{minipage}{14pc}
\includegraphics[width=14pc]{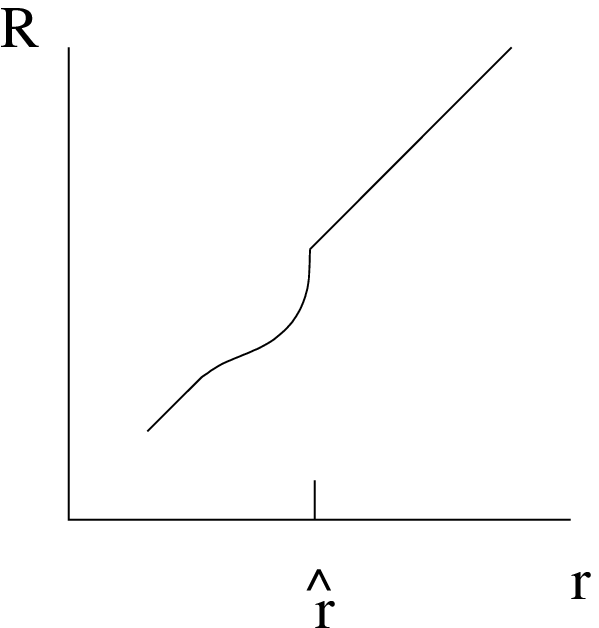}
\caption{\label{outer}Outer gauge}
\end{minipage}\hspace{2pc}
\begin{minipage}{14pc}
\includegraphics[width=14pc]{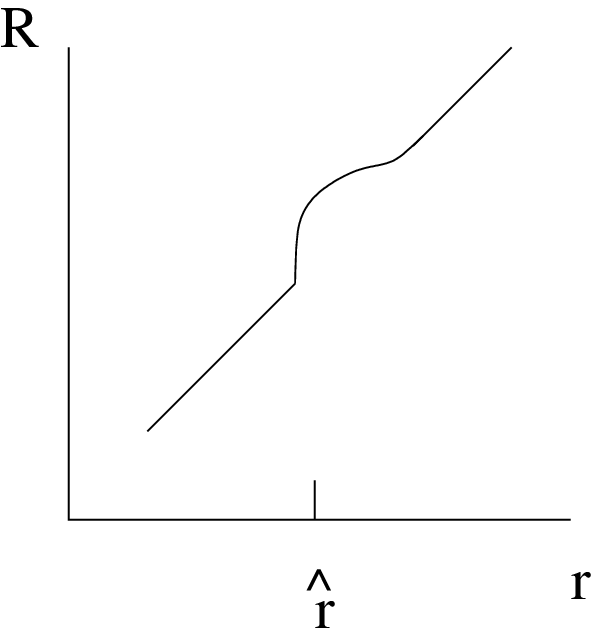}
\caption{\label{inner}Inner gauge}
\end{minipage} 
\end{figure}

\begin{figure}[h]
\includegraphics[width=14pc]{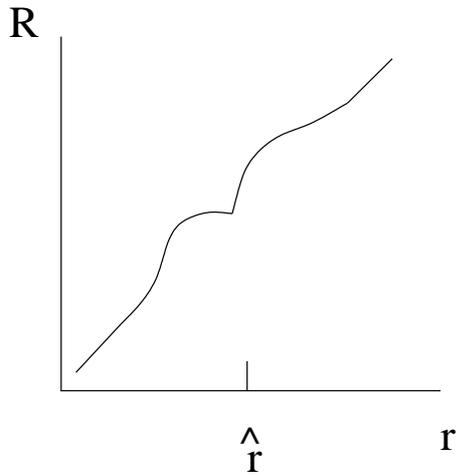}\hspace{2pc}
\begin{minipage}[b]{10pc}\caption{\label{generic}Generic gauge}
\end{minipage}
\end{figure}

Typical are the ``outer gauge'' characterized by $R(r,t)=r$ for $r\geq
\hat r(t)$ and shown in Fig.\ref{outer} and the inner gauge characterized 
by $R(r,t)=r$ for $r\leq \hat r(t)$ shown in Fig.\ref{inner}, but there are
also more general choices as shown in Fig.\ref{generic}. We stress that such
deformation $g$ is not related to the thickness of the shell which is always
zero.  

Using the solutions (\ref{piLfromF},\ref{piRfromF}) of the constraints 
one obtains in the outer gauge
the following reduced \cite{FM} 
action where only the shell position $\hat r$ appears as degree of freedom
\begin{equation}\label{outerreducedaction}
\int_{t_i}^{t_f} \left(p_{c}~ \dot{\hat r} -\dot
M(t)\int_{r_0}^{\hat r(t)}\frac{\partial F}{\partial M} dr+ 
\left.(-N^r \pi_L + NRR')\right|^{r_m}_{r_0}\right)dt.
\end{equation}
The $p_c$ is easily computed
\begin{equation}\label{pcgeneral}
p_c = \hat r (\Delta {\cal L}-\Delta {\cal B})
\end{equation}
where  $\Delta {\cal L}={\cal L}(\hat r+\varepsilon)-{\cal L}(\hat
r-\varepsilon)$ and similarly for $\Delta{\cal B}$. 
In deriving eq.(\ref{pcgeneral}) we used the consequences of the constraints
(\ref{constraintr},\ref{constraintt})
\begin{equation}\label{disceq}
\Delta R' =-\frac{V}{R};~~{\rm where~~} V=\sqrt{\hat
p^2+m^2};~~~~\Delta\pi_L=-\hat p.
\end{equation}
In the outer gauge we find
\begin{equation}\label{pc}
p_c=  \sqrt{2M\,\hat r}-\sqrt{2H \,\hat r}-\hat
r\log\left(\frac{\hat r+\sqrt{{\hat p}^2+m^2}-\hat p-
\sqrt{2H \,\hat r}}{\hat r-\sqrt{2M \,\hat r}}\right)
\end{equation}
with $\hat p$ given implicitly by
\begin{equation}\label{fundamentalH}
H-M= V +\frac{m^2}{2\hat r}-\hat p\sqrt{\frac{2H}{\hat r}};~~~~~
V=\sqrt{{\hat p}^2+m^2}.
\end{equation}
This is the result obtained by Friedman, Louko, Winters-Hilt \cite{FLW}
through a limit procedure in which the support of the deformation function
goes to zero but actually it is
independent of the deformation. For working out the dynamics of one shell the
limit procedure is all right, but in dealing with two or more shell it is
important to have a deformation on a finite range because otherwise a limit
procedure would give a result dependent on the order in
which the two limits, deformation and $\hat r_1\rightarrow \hat r_2$, are
taken.

Similarly one can compute using the general formula the reduced canonical
momentum of the system in the inner gauge
\begin{equation}
p_c^i= \sqrt{2M\,\hat r}-\sqrt{2H \,\hat r}-\hat
r\log\left(\frac{\hat r-\sqrt{2H \,\hat r}}
{\hat r- V + \hat p-\sqrt{2M\hat r}}\right)
\end{equation}
and $\hat p$, again determined by the discontinuity equation
(\ref{disceq}), is given now by the implicit equation
\begin{equation}\label{fundamentalM}
H-M=V-\frac{m^2}{2\hat r}-\hat p \sqrt{\frac{2M}{\hat r}};~~~~V=\sqrt{\hat
p^2+m^2}.
\end{equation}
The two reduced canonical momenta $p_c,~p_c^i$ appear to be completely 
different but
they can be proven to be the same function of $\hat r$.
One can consider also the more general gauges depicted in Fig.3 and 
it can be proven \cite{menotti} that $p_c$ is always the same. 
However in inner gauge a term $\dot H(t)$ appears in the reduced action while 
in the 
more general gauges both a term $\dot M$ and a term $\dot H$ appear in the 
action.

\noindent The boundary term given in eq.(\ref{completeaction}) is 
equivalent to
\begin{equation}
- H N(r_m) + M N(r_0).
\end{equation}

\section{Equations of motion}\label{eqmotionSec}

In deriving the equation of motion from action (\ref{outerreducedaction}) one 
can consider $M(t)=M$, the interior mass, as a  datum of the problem and 
vary $H$ to obtain
\begin{equation}
\dot{\hat r} \frac{\partial p_c}{\partial H} - N(r_m)=0 
\end{equation}
which using the expression of $p_c$ and the relation between $N$ and $N^r$
imposed by the gravitational equations \cite{FM,FLW} can be written as
\begin{equation}\label{1sheqmotion}
\dot{\hat r} = \frac{\hat p}{V}N(\hat r)- N^r(\hat r) =
\left(\frac{\hat p}{V}-\sqrt{\frac{2H}{\hat r}}\right)N(\hat r).
\end{equation}
Alternatively one can consider $H$, the total energy as a datum of the problem
and vary $M(t)$. The calculation is far more complicated due to the presence
of $\dot M$ in the action (\ref{outerreducedaction}), but using the
equations for the gravitational field one reaches \cite{FM} the same equation 
of motion (\ref{1sheqmotion}). In addition a consequence of the gravitational 
equations of motion is the constancy in time of $M(t)$ and $H(t)$.

\section{Analytic properties of $p_c$}\label{analyticSec}

We saw that in the outer gauge $p_c$ is given by (\ref{pc},\ref{fundamentalH}).
The solution of eq.(\ref{fundamentalH}) for $\hat p$ is
\begin{equation}
\frac{\hat p}{\hat r} =\frac{A \sqrt{\frac{2H}{\hat r}}~
\pm\sqrt{A^2 -(1-\frac{2H}{\hat r})\frac{m^2}{\hat r^2}}}
{1-\frac{2H}{\hat r}}
\end{equation}
where
\begin{equation}
A =\frac{H-M}{\hat r}-\frac{m^2}{2 \hat r^2}.
\end{equation}
If we want $\hat p$ to describe an outgoing shell we must choose the
plus sign in front of the square root. Moreover the shell
reaches $r=+\infty$ if and only if $H-M>m$ as expected. 

The logarithm in $p_c$, eq.(\ref{pc}), has branch points at zero and infinity 
and thus we must
investigate for which values or $\hat r$ such values are reached.
At $\hat r=2H$, $\hat p$ has a simple pole with positive residue. Thus the
numerator in the argument of the logarithm in $p_c$ goes 
to zero and below $2H$ it becomes
\begin{equation}\label{contnumerator}
\hat r - V - \hat p -\sqrt{2H\hat r}
\end{equation}
where here $V$ is the absolute value of the square root. Expression 
(\ref{contnumerator}) is negative irrespective of the sign of $\hat p$ and
stays so for $\hat r<2H$
because $\hat p$ is no longer singular.
As a consequence $p_c$
below $2H$ acquires the imaginary part $i\pi \hat r$. 
Below $\hat r =
2M$ the denominator of the argument of the logarithm in 
eq.(\ref{pc})
becomes negative so that the argument of the logarithm reverts to positive
values. 
Thus the so called classically forbidden  region where $p_c$ becomes complex 
is $2M<\hat r <2H$
independent of $m$ and of the deformation $g$ and the integral of the 
imaginary part of $p_c$ for any deformation $g$ and for any mass $m$ of the
shell is
\begin{equation}\label{integratedimpart}
{\rm Im}~\int p_c dr = 
\pi \int_{2M}^{2H} r dr = 2\pi(H^2-M^2)= \frac{\Delta S}{2}
\end{equation}
which is the original result derived in \cite{KW1,PW} 
in the zero mass case. Parikh and Wilczek \cite{PW} gave to 
\begin{equation}
\exp (-2 {\rm Im}\int p_c dr )
\end{equation}
the interpretation of the tunneling 
probability for the emission of a quantum of energy $\omega$. Criticism and
alternative  proposals for the emission probability like 
\begin{equation}
\exp (- {\rm Im} \oint p_cdr )
\end{equation}
\begin{equation}
\exp \left(-2 {\rm Im}(\int p_c dr +  {\rm temporal~contribution} )\right)
\end{equation}
followed \cite{chowdhury,zerbini,akhmedov,akhmedova,
zerbini1,zerbini2,zerbini3,pizzi,belinski,padmana,banerjee,kerner}. 
Here instead we shall discuss the
role of eq.(\ref{pc}) in the framework of  mode analysis. This will be done in
Sec.(\ref{modeSec}).

\section{Two shell reduced action}\label{twoshellSec}

\begin{figure}
\begin{center}
\includegraphics{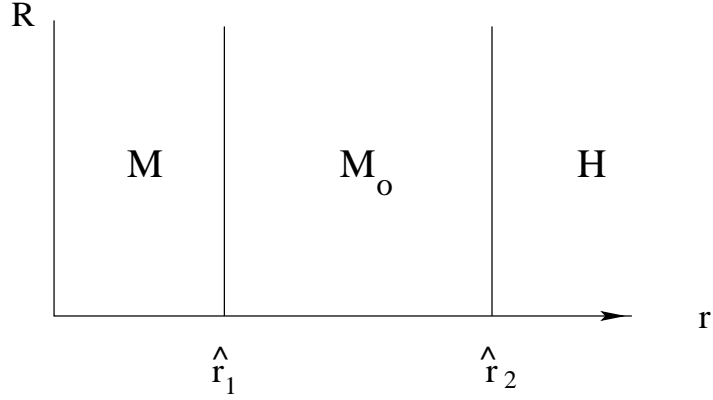}
\end{center}
\caption{\label{2sh} Two shell dynamics}
\end{figure}

Now we have three characteristic masses, $M$, $H$ and an intermediate mass
$M_0$, see Fig.4, which can change only if the two shells cross. Working as 
before with $M={\rm const}$ considered as a datum of the
problem we reach the reduced action \cite{FM}
\begin{eqnarray}\label{twoshellreducedaction}
&& \dot{\hat r}_1p_{c1} + \dot{\hat
r}_2p_{c2}+ \dot H (R(\hat r_1)-\hat r_1)\frac{\partial T}{\partial
H}{\cal D} + \dot M_0 (R(\hat r_1)-\hat r_1)\frac{\partial T}{\partial M_0}
{\cal D}+ \nonumber\\
&& +\frac{d}{dt}\int_{r_0}^{\hat r_2} F dr-\dot M_0\int^{\hat r_2}_{\hat
r_1}\frac{\partial F}{\partial M_0}dr 
+(-N^r\pi_L+N R R')|^{r_m}_{r_0} 
\end{eqnarray}
where
\begin{eqnarray}\label{pc1}
T=\log \frac{V_2}{R(\hat r_2)};~~~~
{\cal D} = R (\Delta{\cal L}- \Delta{\cal B})|_{\hat r_1};\\
p_{c1}= R'(\hat r_1+\varepsilon){\cal D};~~~~
p_{c2}= p_{c2}^0+\frac{d}{d\hat r_2}(R(\hat r_1)-\hat r_1){\cal D}
\end{eqnarray}
and $p^0_{c2}$ is given by eq.(\ref{pc}) with $M$ replaced by $M_0$.
The novelty is that now even in the outer gauge the time derivative of $H$
intervenes in addition to $\dot M_0$ and $p_{c1}$ and $p_{c2}$ depend both on
$\hat r_1$ and $\hat r_2$. We can vary $\hat r_1$, $\hat r_2$, $H$ and $M_0$ 
independently
obtaining the correct equations of motion \cite{FM}. As expected one finds 
that 
the exterior shell moves irrespective of the dynamics which develops at lower
values of $r$ until a crossing occurs.

\section{The exchange relations}\label{exchangeSec}

In case of crossing of the two shells from the equations of
Sec.(\ref{twoshellSec})  we can obtain relations between $M, M_0,H, \hat r_e$ 
and
$M'_0$ being $\hat r_e$ the the shell position at the crossing and $M'_0$ the
intermediate mass after the crossing. During the crossing the masses of the 
shells can change, provided they satisfy a relation analogous to the 
energy-momentum conservation is special relativity
\begin{eqnarray}
& & \hat p_1+\hat p_2=\hat p'_1+\hat p'_2\\
& & V_1+V_2=V'_1+V'_2;~~~~V_n = \sqrt{\hat p_n^2+m_n^2};~~~~V'_n = \sqrt{\hat
{p'}_n^2+{m'}_n^2}. \nonumber
\end{eqnarray}
This is an outcome of the constraints. 
It is not possible to predict the final masses of the two shells as it depends
on the details of the interaction which has to be specified. 
A relatively simple case is when the masses are unchanged during the crossing
(transparent crossing) \cite{FM} and the simplest case is 
the crossing of two massless shells
which remain massless, thus re-obtaining the well known Dray-'t Hooft-Redmount
relations \cite{DtH,redmount} which we report here below
\begin{equation}
H \hat r_e + M \hat r_e-2 H M = M_0 \hat r_e- 2 M_0 M'_0 + M'_0 \hat r_e
\end{equation}
being $\hat r_e$ the crossing radius.
\begin{figure}
\begin{center}
\includegraphics{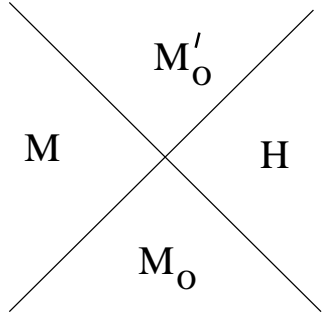}
\end{center}
\caption{\label{DHR}The exchange diagram}
\end{figure}

\section{Integrability of the form $p_{c1}d\hat r_1+p_{c2}d\hat r_2$ and
imaginary part of the space- component of the action} 
\label{integrabilitySec}

The space-part of the on-shell action for two massive shells of matter is 
given from eq.(\ref{twoshellreducedaction}) by
\begin{equation}
\int_{t_i}^{t_f} 
(p_{c1}~\dot{\hat r}_1+ p_{c2}~\dot{\hat r}_2)~dt
\end{equation}
as on the equations of motion $\dot H(t)=\dot M(t)=0$.
It is possible to prove a theorem analogous to the one found in the books of
Whittaker and  Arnold \cite{whittaker,arnold} i.e. that in presence of a 
constant 
of motion, in addition to the hamiltonian, the form $p_{c1} d\hat r_1+
p_{c2} d\hat r_2$ is 
closed even though the proof is somewhat different \cite{FM}. The intermediate 
mass $M_0$ plays the role of the additional constant of motion. 
The above result allows to deform the integration path as to bring $\hat
r_1$  immediately below or above $\hat r_2$. This is allowed by the absence of
discontinuities in our scheme. But, in words, two coaleshed shell have the 
same properties of a single shell with energy $H-M$. The final result is that
\begin{equation}\label{intpc1pc2}
{\rm Im}\int_{t_i}^{t_f} dt 
(p_{c1}~\dot{\hat r}_1+ p_{c2}~\dot{\hat r}_{c2})=
{\rm Im} \int_{r_{1i}, r_{2i}}^{r_{1f}, r_{2f}} 
(p_{c1}~d\hat r_1+p_{c2}~d\hat r_2) = 2\pi(H^2-M^2).
\end{equation}
The result (\ref{intpc1pc2}) holds also for in the case when at the crossing
the two shells can change their mass. For details we refer to \cite{FM}.
 
\section{Mode analysis}\label{modeSec}

The original way to extract information on the spectrum of the radiation is
mode analysis \cite{hawking,KW1,KVK}. In the massless case
\begin{equation} \int^{\hat r} p_c d\hat
r' = f(\hat r, M)-f(\hat r, H) 
\end{equation} can be computed exactly and its
expansion to first order in $\omega=H-M$ is 
\begin{equation} \int^{\hat r} p_c
d\hat r'= 4 M \omega\log(\hat r - 2M)+~{\rm regular~terms}. 
\end{equation}
Given the semiclassical mode 
\begin{equation} \phi(r,t)=e^{i\int^{\hat r} p_c d\hat r' -
i\omega t} \end{equation} we can perform the analysis of the mode regular at
the horizon in terms of the above mode. This analysis can be
performed either by scalar product \cite{BD} i.e. space integration 
where one has to keep into account that the background metric is Painlev\'e 
or by time Fourier
analysis. In this way one obtains the well known formulas for the Bogoliubov
coefficients. The regular outgoing modes near the horizon are given by
\begin{equation} 
\psi(\hat r,t)=e^{ik(\hat r-2M)e^{-\frac{t}{4M}}}. 
\end{equation}
Computing the scalar product, and taking into account that the background
metric is the Painlev\'e metric, we have 
\begin{equation}\label{scalarproduct} 
-i\int(\psi^*\partial_\rho\phi-\phi\partial_\rho\psi^*)g^{\rho
0}\varepsilon_{0r\theta\phi}\sqrt{-g} ~d\hat r d\theta d\phi 
\end{equation} 
with
\begin{equation} 
g^{rt}=N^r=\sqrt{\frac{2M}{\hat r}}~~~~{\rm and}~~~~\sqrt{-g}=1
\end{equation} 
and the integration region is outside the horizon. 
Keeping only the most singular terms at the horizon eq.(\ref{scalarproduct}) 
reduces, with $\tau=\exp(-t/4M)$,  to
\begin{equation} 
\int_0^\infty e^{-ikx\tau} e^{4iM\omega \log x -i \omega t}
~\frac{dx}{x}. 
\end{equation} 
We can compute such scalar products at $t=0$ obtaining the standard 
Hawking integral 
\begin{equation}\label{alphakomega} 
\int_0^{\infty}e^{-ikx} e^{4iM\omega \log
x} ~\frac{dx}{x}= e^{2\pi\omega M}(k)^{-4i\omega M}\Gamma(4i\omega M)= {\rm
const}~\alpha^*_{k\omega} 
\end{equation} which gives the dominant contribution for large
$k$. The coefficient $\beta_{k\omega}$ is obtained by changing in
(\ref{alphakomega}) $\omega$ into $-\omega$. 
Alternatively we can extract the Bogoliubov coefficients by performing a
time Fourier transform i.e. 
\begin{equation} 
\int_{-\infty}^{+\infty}e^{-ikx\tau} e^{4iM\omega\log x -i\omega t}dt=
\int_0^{+\infty} e^{-ikx\tau} e^{4iM\omega
\log(x\tau)} ~\frac{d\tau}{\tau} 
\end{equation} 
which is, as it should be, a
result independent of $r$ (i.e. $x$) and reproduces eq.(\ref{alphakomega}). 
The derivation is valid 
to first order in $\omega$ which is the realm of validity of the external field
approximation. 
For finite $\omega$ it is
problematic to perform a space integration on modes, because we have not a
well defined background metric.

\noindent Thus Kraus and Wilczek \cite{KW1} followed the time Fourier analysis
method. In order to
do so one has to construct the non-perturbative modes regular at the horizon.
This is not a completely trivial task.
The regular modes are given by 
\begin{equation} 
e^{iS}=e^{i k \hat r(0) + i \int_{\hat r(0)}^r p_c d\hat r - 
i(H-M) t}
\end{equation} 
where $S$ is the on shell action computed with the following boundary
conditions: 1) At 
time $t$ the shell position is $\hat r$, outside the horizon. 
2) At time $t=0$ the conjugate
momentum is $k$, i.e. $S(0,\hat r)=k\hat r$. 
Due to these boundary conditions $\hat r(0)$, $H$ and as a consequence
$p_c$ depend on $k,t,\hat r$ even if along each trajectory $H$ is always a
constant 
of motion.
Using the saddle point approximation 
\cite{KW1,KVK}  
one obtains that the absolute value of the Bogoliubov  coefficient
$\alpha_{k\omega}$ is given by
\begin{equation}\label{kvkintegral} 
\left|e^{i\int_{r(0)}^r p_c(r',H, M) dr'}\right|= e^{-{\rm
Im}\int_{r(0)}^r p_c(r',H, M) dr'} 
\end{equation}  
computed for $H=M+\omega$. What is important here is that only the 
space part of the action appears, as the time part 
$(H-M)t$ cancels 
with $\omega t$ at the saddle point. 
$\hat r(0)$ is given by the condition 
\begin{equation} 
k = \hat r(0) \log\frac{\sqrt{\hat r(0)}-\sqrt{2M}}
{\sqrt{\hat r(0)}-\sqrt{2H}}
\end{equation} 
which for $H=M+\omega>M$ is solved by
\begin{equation} 
2M<2H<\hat r(0)
\end{equation} 
and thus there is no imaginary contribution to the integral from the gap
$2M,2H$ as we discussed in Sect.(\ref{analyticSec}).
The Bogoliubov coefficient $\beta_{k\omega}$ is obtained by changing in 
(\ref{kvkintegral}) $\omega$ in $-\omega$ (always $\omega>0$). 
Then we have
\begin{equation} 
\hat r(0)<2H<2M
\end{equation} 
The value of the saddle point $t$, now complex, is given by
\begin{equation} 
t = 4H\log\frac{\sqrt{\hat r}-\sqrt{2H}}{\sqrt{\hat r(0)}-\sqrt{2H}}
\end{equation} 
and the r.h.s. term of eq.(\ref{kvkintegral}) becomes
\begin{equation}\label{correctedbeta}
e^{-4\pi M \omega(1 -\omega/2 M)}
\end{equation}
where the last passage is due to eq.(\ref{integratedimpart}) for
which we gave a general proof within the family of Painlev\'e gauges even in
the massive case.
An explicit derivation of the result (\ref{correctedbeta}) has been given 
in \cite{menotti}
by working out the late time expansion of the time development of the action. 
The main point is that in the mode analysis what comes in is not 
the total
action of a  model particle crossing the horizon, but only the ``space
part'' of it. The analysis and the results depend on the interpretation of the
expression 
\begin{equation} 
\exp{(iS)}
\end{equation} 
as modes of the field dressed by the gravitational interaction with non 
vanishing quanta of energy.
The result eq.(\ref{intpc1pc2}) proven for the emission of two 
shells \cite{FM} was
interpreted \cite{parikh} as the absence of correlations among quanta in the
emitted radiation. It would be of interest to give a similar ``mode
interpretation'' of the result (\ref{intpc1pc2}) derived for the
emission of 
two shells which during the time evolution can also interact \cite{FM}. 
To this end one should compute the
two shell modes which are regular at the horizon, and perform a time Fourier
analysis of them. This has not yet been accomplished.

\section{Concluding remarks}\label{conclusionSec}

In this paper we gave a general treatment of the dynamics of one or more self-
gravitating spherical shell of matter in a spherical gravitational field. We
extended the treatment of \cite{KW1} to more than one shell \cite {FM} proving
on the way the universality of the reduced conjugate momentum within the
family of Painlev\'e gauges \cite{menotti}.

This allows to give a general derivation of the exchange relations and also,
exploiting an integrability result, to extend the result on the imaginary part
of the space part of the action to more than one shell. Instead of following
the tunneling picture here we have revisited the treatment which follows by
interpreting the exponential of the action, properly subtracted, as the
dressed modes of the Hawking radiation. The main point here is that according
such a mode interpretation only the space part of the action and in particular
its imaginary part comes in determining the Bogoliubov coefficient and thus
in determining the Hawking spectrum corrected for the back reaction effects.

\end{document}